\documentstyle[emulateapj]{article}
\lefthead{Hughes}
\righthead{X-Ray Expansion of Tycho's SNR}
\submitted{Received 2000 September 6; accepted 2000 October 6}
\accepted{To Appear in the Astrophysical Journal Letters}
\def\einstein{{\it Einstein}}

\def\rosat{{\it ROSAT}}
\def\asca{{\it ASCA}}

\def\myarcsec{\mskip1mu^{\prime\prime}\mskip-7mu.\mskip2mu}
\def\lsim{\hbox{\raise.35ex\rlap{$<$}\lower.6ex\hbox{$\sim$}\ }}
\def\gsim{\hbox{\raise.35ex\rlap{$>$}\lower.6ex\hbox{$\sim$}\ }}

\begin{document}

\title{The Expansion of the X-ray Remnant of Tycho's Supernova
(SN1572)}

\author{John P. Hughes}
\affil{Department of Physics and Astronomy, Rutgers
University, 136 Frelinghuysen Road, Piscataway, NJ 08854-8019
jph@physics.rutgers.edu}

\begin{abstract}

Two \rosat\ high resolution images separated by nearly five years have
been used to determine the expansion of the X-ray remnant of Tycho's
supernova (SN1572). The current expansion rate averaged over the
entire remnant is $0.124\pm0.011 \,\% \,\rm yr^{-1}$, which, when
combined with the known age of the remnant, determines the mean
expansion parameter $m$, defined as $R\propto t^m$, to be
$0.54\pm0.05$. There are significant radial and azimuthal variations
of the X-ray expansion rate. The radial expansion in particular shows
highly significant evidence for the more rapid expansion of the
forward blast wave as compared to the reverse-shocked ejecta, an
effect that has not been seen previously.  The expansion parameter
varies from $m=0.71\pm0.06$ at the outermost edge of Tycho's supernova
remnant (SNR) to a value of $m=0.34\pm0.10$ on the inside edge of the
bright rim of emission. These values are consistent with the rates
expected for a remnant with constant density ejecta evolving into a
uniform interstellar medium during the ejecta-dominated phase of
evolution.  Based on the size, age, and X-ray expansion rates, I
obtain values for the explosion energy and ambient density of
$E\approx 4-5\times 10^{50}\,\rm ergs$ and $n_0 \approx 0.35-0.45\,
\rm cm^{-3}$.  As is also the case for Cas A and Kepler's SNR, the
X-ray expansion rate of Tycho's SNR appears to be significantly higher
than the radio expansion rate.  In the case of Tycho's SNR, however,
the difference between radio and X-ray expansion rates is clearly
associated with the motion of the forward shock.

\end{abstract}


\section{Introduction}

Measurement of the current rate of expansion provides essential
information on the dynamical state of supernova remnants (SNRs). This
is particularly true for the so-called historical remnants, those for
which the date of the supernova (SN) explosion is known, since a
comparison of the average expansion rate with the current expansion
rate yields a measure of the deceleration.  Unfortunately, the rate of
expansion, although rapid compared to most astrophysical objects in
the cosmos, is still rather long on human timescales, necessitating
measurements over the course of years, if not decades, in order to
attain accurate results.  \rosat\ was the first X-ray observatory that
had sufficient angular resolution ($\sim$4$^{\prime\prime}$ half-power
radius for the High Resolution Imager) and operated for a long enough
time that significant measurements of the expansion rate of young SNRs
are possible.

\par

The time-averaged expansion rate of Tycho's SNR based on the outermost
extent of the remnant ($\sim$8$^\prime$ in diameter) in either the
radio or X-ray band and its well-known age is $\sim$0\farcs56
yr$^{-1}$.  The observed proper motion of the optical filaments, which
are believed to trace the location of the blast wave, indicate much
lower current expansion rates ranging from 0\farcs18 yr$^{-1}$ to
0\farcs28 yr$^{-1}$ (Kamper \& van den Bergh 1978).  Evidently these
filaments are locations where significant deceleration of the SN blast
wave has occurred.  From the width of the broad H$\alpha$ emission
Smith et al.~(1991) derive shock velocities in the range 1500--2800 km
s$^{-1}$ that, when combined with the proper motion measurements,
imply a distance of 1.5--3.1 kpc.  This range is in good agreement
with other distance estimates to the remnant (Green 1984; however, see
Schwarz et al.~1995) and in the following I adopt a value of 2.3 kpc
for the distance.  Because the optical filaments cover only a limited
portion of Tycho's SNR, the optical data are unable to provide a
comprehensive picture of the expansion of the remnant.


\par

The radio remnant of Tycho's SN has also been observed to be expanding
at the current epoch (Strom, Goss, \& Shaver 1982; Tan \& Gull 1985;
Reynoso et al.~1997, hereafter R97). Although in each successive
analysis the radio data have improved, the basic result of these
studies have remained in general agreement. The current radio
expansion rate, averaged over the outer rim, is 0.113 \% yr$^{-1}$ or,
expressed equivalently, the expansion parameter, defined as $R\propto
t^m$, is $m=0.471\pm0.028$.  This result is between the free expansion
rate, $m\sim 1$, and the expansion rate expected for a remnant in the
Sedov phase of evolution, $m=0.4$. There is significant azimuthal
variation of the radio expansion rate, while interior features appear
to show the same expansion rate as the rim (R97).


Vancura, Gorenstein, \& Hughes (1995), using data from two satellite
observatories, quoted a current X-ray expansion rate for the SNR based
on a 11.5 yr time baseline that was consistent with these other
values. Here I present more accurate results on the X-ray expansion of
Tycho's SNR using high resolution images accumulated by the \rosat\
satellite in two epochs separated by 4.55 yr.  A preliminary report on
this work, using a different analysis approach, was given by Hughes
(1997).

\section{Observations}

The reduction of the X-ray data closely follows that done in an
earlier study of Kepler's SNR (Hughes 1999, hereafter H99); interested
readers are referred there for more details.  A log of the high
resolution \einstein\footnotemark\footnotetext{
The \einstein\ observation is included here for completeness and in
fact results for only the two \rosat\ data sets are presented below.
Preliminary studies found that the \einstein/\rosat\ expansion results
were inconsistent with the \rosat/\rosat\ ones, yielding expansion
values a factor of 1.6--1.9 times higher.  The \einstein\ and \rosat\
comparison is subject to more uncertainty due to differences in the
instrumental point-spread functions and spectral bandpasses.  The
latter effect, combined with spatial variations in the X-ray spectrum
of Tycho's SNR (Vancura et al.~1995), is the likely cause of the
discrepancy.
}
and \rosat\ imaging observations of Tycho's SNR
is given in Table 1.  The columns list the observatory, start date,
the Modified Julian Day (MJD) corresponding to the average date of the
observation, and the effective duration (live-time corrected). Figure
1 shows the image from observation R2.

\begin{small}
\begin{minipage}[t]{83mm}
\begin{center}
{\noindent{TABLE 1}}\\
{\noindent{\sc Observations of Tycho's SNR}}\\[6pt]
\begin{tabular}{@{}lccc@{}}
\hline\hline\\[-4pt]
Observatory
& Start Date
& Ave.~MJD
& Duration (s)\\[4pt]
\hline\\[-4pt]
{\it Einstein}      & 1979 Feb 8  & 43,913.3 & $\phantom{0}50409.2$\\[2pt]
{\it ROSAT}    (R1) & 1990 Jul 28 & 48,100.7 & $\phantom{0}22163.7$\\[2pt]
{\it ROSAT}    (R2) & 1995 Feb 5  & 49,763.7 & $104332.1$\\[4pt]
\hline
\end{tabular}
\end{center}
\end{minipage}
\end{small}

\begin{center}
\begin{minipage}[t]{0.47\textwidth}
\vspace{-0.5in}
\epsfxsize=1.1\textwidth \epsfbox{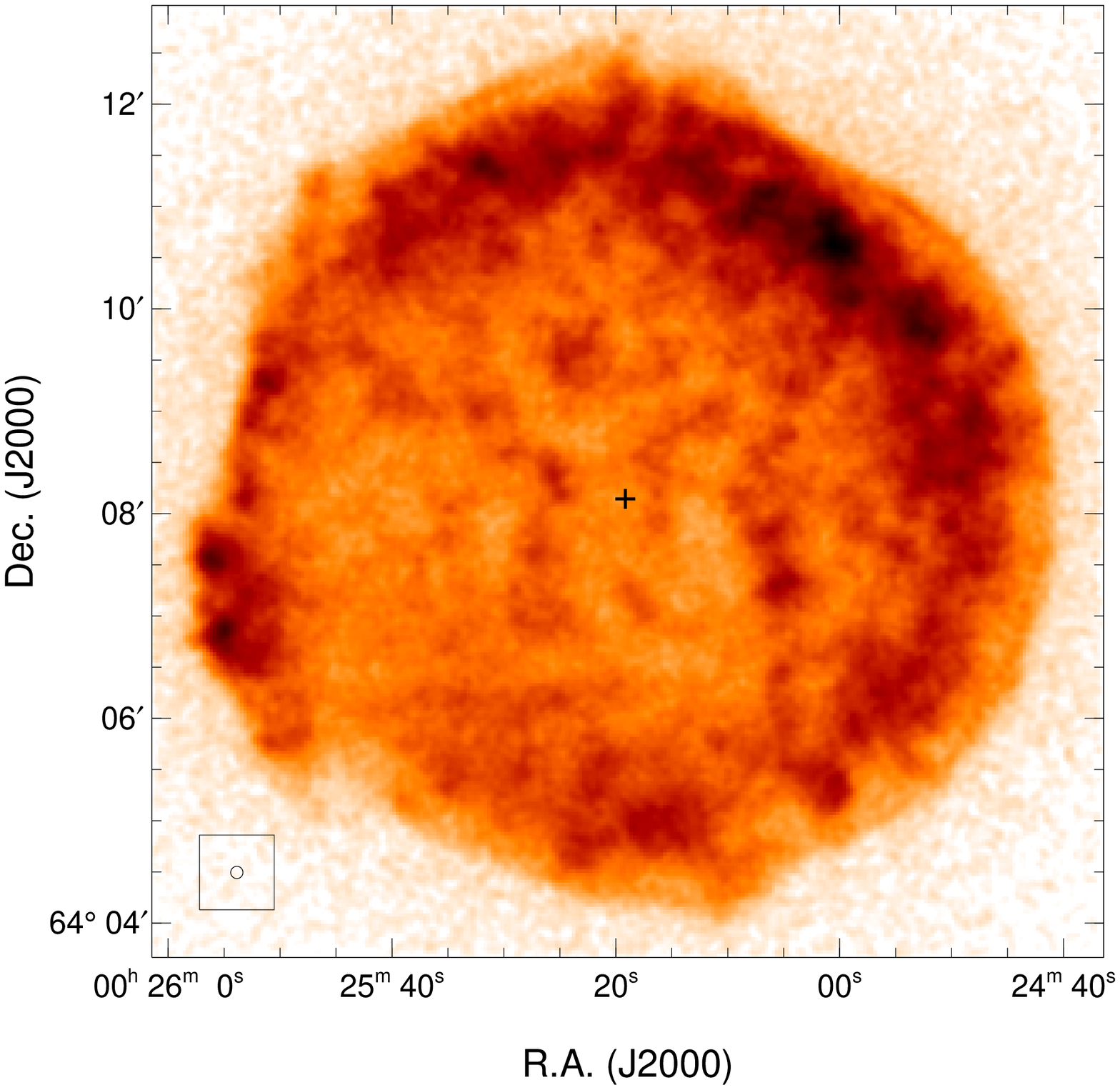}
\figcaption{
{\it ROSAT\/} HRI image of Tycho's SNR.  The data were smoothed by a
Gaussian function with $\sigma = 2^{\prime\prime}$ and are displayed
with a square-root intensity scaling. The plus sign marks the
geometric center of the remnant. The effective resolution of the map,
including both the instrumental point-spread function and the average
width of the smoothing kernel, is shown at the lower left.
}
\end{minipage}
\end{center}

The \rosat\ high resolution imager (RHRI) data were processed in some
detail.  The data were filtered in pulse height to reduce background.
Pulse height channels 1 to 11 were used for the first epoch \rosat\
observation (R1) and channels 1 to 12 were used for the second epoch
observation (R2).  This reduced the background level by 5\%-7\% while
the source rates were nearly unaffected ($\sim$1\% change).  Aspect
drift throughout an observation was corrected by aligning separate
images made from the data corresponding to each orbit.  The images
from all the sub-intervals (typically 1500 s long) were registered to
the nearest $0\myarcsec5$ pixel, shifted, and added.  For R1 the
initial registration of the individual maps from the standard analysis
was fairly good: all of the individual maps were already aligned
relative to each other to within $\sim$2$^{\prime\prime}$ or better.
In the case of R2 there was clear evidence for a drift in aspect
throughout the observation.  The mean registration error was
$\sim$3$^{\prime\prime}$, although some of the individual images were
misregistered by up to 10$^{\prime\prime}$.  For both epochs, the
shift-and-add alignment technique produced images with a noticeably
improved point response function.

The grain scattering halo from Tycho's SNR (Mauche \& Gorenstein 1986;
Predehl \& Schmitt 1995) extends over the entire field of view of the
HRI, which makes background estimation difficult.  The background
level was estimated by fitting a spatial power-law component (for the
scattering halo) plus a constant background level to the surface
brightness profile over the 5$^\prime$ to 15$^\prime$ radial
range. The fitted power-law components were consistent between the two
RHRI pointings (index of $-$2.3), although the background levels
differed by some 15\%, $4.2\times 10^{-3}$ cts s$^{-1}$ arcmin$^{-2}$
for R1 and $3.7\times 10^{-3}$ cts s$^{-1}$ arcmin$^{-2}$ for R2.
This difference is within the variation observed from field to field
for the RHRI (David et al.~1998).

Exposure maps were generated for the RHRI observations as before (see
H99). Over the portion of the field containing the image of the
remnant the ratio of exposure between the first and second epochs
varied between 0.963 and 1.045. These corrections are small in
comparison to the flux differences I measure and are uncorrelated with
image structure in the remnant. Nevertheless, these maps are included
in the model fits described below.

The exposure- and deadtime-corrected, background-subtracted RHRI count
rates of Tycho's SNR within a radius of 8$^\prime$ are $9.092\pm
0.020\,\rm s^{-1}$ and $9.209\pm 0.009\,\rm s^{-1}$ from the first and
second epoch images, respectively.  This difference in count rates
($\sim$1\%) is consistent with the range of RHRI rates seen in
calibration observations of the SNR N132D (see H99).

Finally I comment on possible changes in the plate scale using
observations of the Andromeda galaxy, which was observed by the RHRI
in July 1990, 1994, and 1995, and January 1996.  I extracted these
data and used the positions of 10 isolated, moderately bright X-ray
point sources to constrain the relative rotation angle and plate scale
change between pairs of observations. I find no evidence for a change
in the RHRI plate scale and set a limit of $\sim$0.008\% yr$^{-1}$ on
any changes for timescales of 4 yrs or more.

\vspace{-0.25in}

\section{Expansion Results}

The expansion rate was determined using fitting software that takes one
image (``model'') and compares it to another (``data'') as described
in H99.  The model image was scaled in intensity, shifted in position,
and expanded (or contracted) in spatial scale to match the data using,
as the figure-of-merit function, the maximum likelihood statistic for
Poisson-distributed data. The fitted spatial scale factor yields the
global mean expansion rate, which is assumed to be uniform across the
entire remnant.

Over the 4.55 yrs between observations R1 and R2, Tycho's SNR expanded
annually by an amount $0.124\pm0.011\,\% \,\rm yr^{-1}$.  The error
bar is statistical at $1\,\sigma$ and includes uncertainty from Poisson
noise in both observations as well as plate scale changes assuming the
limit given above. This result is highly significant both in terms of
the final error bar and the reduction in the value of the likelihood
statistic for fits with and without any expansion.

In order to determine the expansion rate as a function of radius and
azimuthal angle, one needs to know the position of the expansion
center.  The current X-ray data do not allow for the determination of
this quantity (see H99), so I have opted to just define the geometric
center of the remnant at $0^{\rm h}25^{\rm m}19^{\rm s}\,\,
64^{\circ}08^{\prime}10^{\prime\prime}$ (J2000) as the nominal center
of expansion. I then investigated how the results depended on this
choice by varying the center position by 40$^{\prime\prime}$ in each
of the four cardinal directions.  I found that the different choices
of center could be nearly perfectly compensated for by appropriate
choices of the relative alignment of the two images. (Since there were
no serendipitous point sources in the field, it was not possible to do
an independent registration of the images.)  This result is not
surprising since non-optimal image alignment or choice of expansion
center will each introduce a sinusoidal term in the expansion rate as
a function of azimuth. In effect what I have done is to remove any
such sinusoidal term from the results, regardless of origin.  My
approach is different from that of R97 who used a
fixed expansion center (defined by the center of the nearly circular
western limb) to measure the radio expansion of Tycho's SNR. And
indeed their fractional expansion results do contain a significant
sinusoidal term.  Because of these complications a detailed comparison
between the X-ray and published radio azimuthal variation is not
particularly enlightening and will not be pursued in this work.
However, since a sinusoid averages to zero over a full cycle, this
difference does not affect the comparison of the radio and X-ray
global mean expansion rates.  As concerns the X-ray azimuthal
expansion, I note that the weighted average rate, $\sim$0.13\%
yr$^{-1}$, is consistent with the global mean X-ray rate and that
there are statistically significant azimuthal variations on angular
scales of 10$^\circ$ to 90$^\circ$ (similar to those in Hughes 1997).

\begin{small}
\begin{minipage}[t]{83mm}
\begin{center}
{\noindent{TABLE 2}}\\
{\noindent{\sc X-ray Expansion Rates with Radius for}}\\
{\noindent{\sc Tycho's SNR}}\\[6pt]
\begin{tabular}{@{}ccc@{}}
\hline\hline\\[-4pt]
Radial range & Exp Rate       & Sys Err  \\
(arcmin)     & (\% yr$^{-1}$) & (\% yr$^{-1}$) \\[4pt]
\hline\\[-4pt]
 0.0 -- 1.5 & $0.212\pm 0.088$ & ($-$0.090, $+$0.036)  \\[2pt]
 1.5 -- 2.0 & $0.178\pm 0.057$ & ($-$0.019, $+$0.008)  \\[2pt]
 2.0 -- 2.4 & $0.124\pm 0.051$ & ($-$0.017, $+$0.016)  \\[2pt]
 2.4 -- 2.8 & $0.133\pm 0.033$ & ($-$0.008, $+$0.006)  \\[2pt]
 2.8 -- 3.2 & $0.080\pm 0.022$ & ($-$0.000, $+$0.006)  \\[2pt]
 3.2 -- 3.6 & $0.107\pm 0.015$ & ($-$0.003, $+$0.005)  \\[2pt]
 3.6 -- 4.0 & $0.117\pm 0.012$ & ($-$0.004, $+$0.006)  \\[2pt]
 4.0 -- 4.4 & $0.167\pm 0.012$ & ($-$0.010, $+$0.008)  \\[2pt]
 4.4 -- 5.2 & $0.176\pm 0.043$ & ($-$0.004, $+$0.003)  \\[4pt]
\hline
\end{tabular}
\end{center}
\end{minipage}
\end{small}

In the remainder of this section I focus on the radial variation of
the X-ray expansion rate.  Here the fits were carried out separately
for several different annular regions about the geometric center of
the remnant.  Most of the radial bins were 24$^{\prime\prime}$ wide
although the innermost and outermost annuli were somewhat thicker.
There were only two fit parameters: the fractional expansion rate and
the change in intensity. Numerical values for the expansion rate are
given in Table 2 and are plotted in Figure 2 along with the change in
the X-ray flux and for reference the surface brightness profile.  In
the bottom two panels the error bars show the statistical
uncertainties, while the small boxes surrounding the data points show
the range of values that come from fits using the four different
expansion centers.  Only for the data point closest to the remnant's
center is this a significant error.  Between the two epochs the X-ray
flux appears to have changed only very slightly over the image of the
remnant, i.e., by less than 2\%.  The weighted average expansion rate
is $\sim$0.13\% yr$^{-1}$, although I can reject the hypothesis that
the expansion rate is constant with radius at more than the 99.9\%
confidence level ($\chi^2 = 17.65$ for 3 d.o.f.) based on the four
data points near the rim (radii between 2\farcm8 and 4\farcm4).  I
find that the expansion parameter increases from $m = 0.34\pm0.10$
just inside the bright rim of Tycho's SNR to a value $m = 0.71\pm0.06$
at the outermost edge of the SNR. The rapid motion of the outermost
edge is the principal cause of the non-uniform radial expansion rate.
In fact, over the interior portion of the SNR, covering radii from
$2\farcm8$ to $4\farcm0$, the X-ray data are consistent with an
expansion parameter of $m=0.45 \pm0.04$.  The large expansion rate for
the outer rim, plus evidence for a slower rate further in, was also
found by Hughes (1997) using an entirely different analysis technique.

\begin{center}
\begin{minipage}[t]{0.47\textwidth}
\epsfxsize=0.98\textwidth \epsfbox{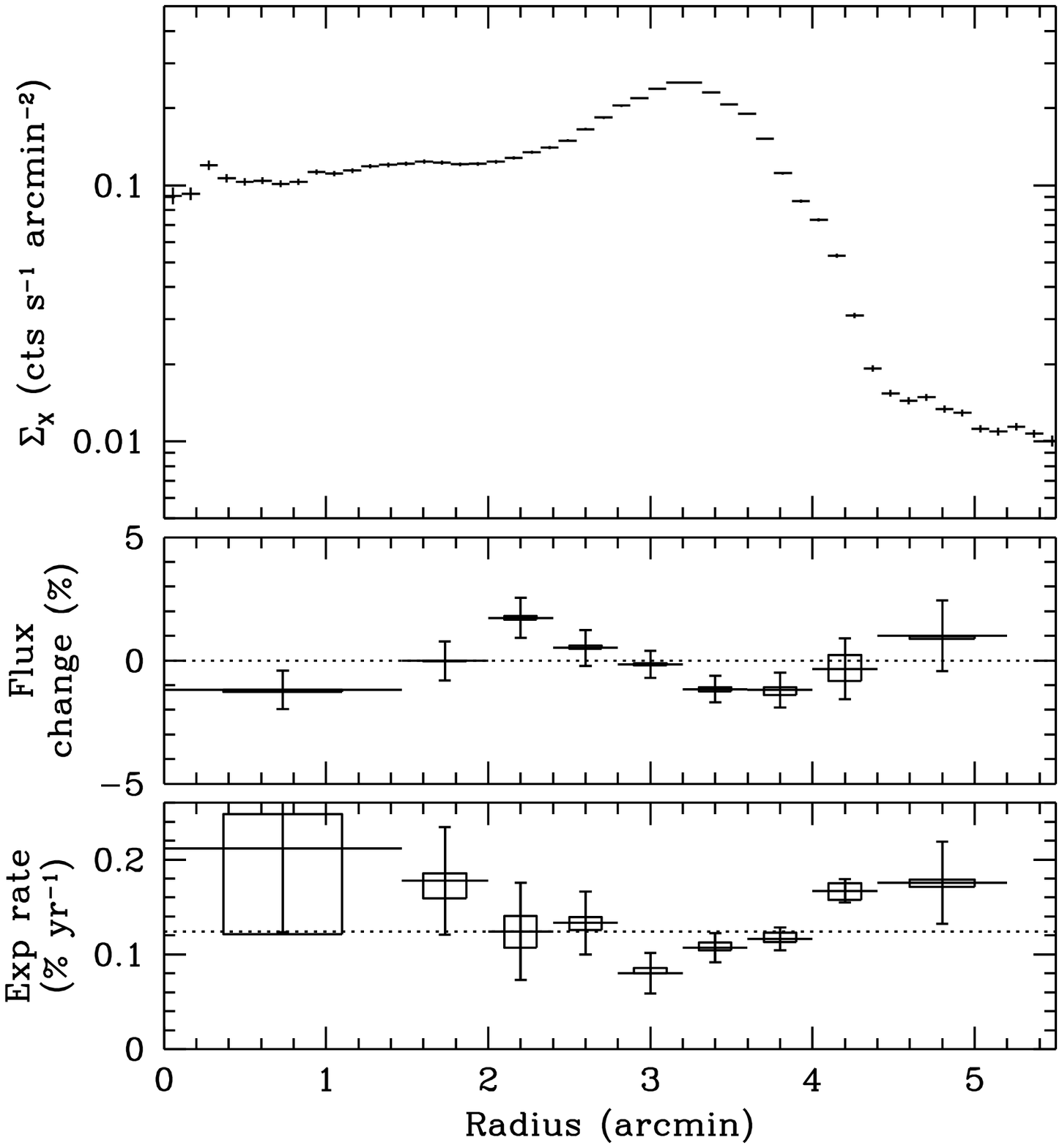}
\figcaption{Radial X-ray surface brightness profile of
Tycho's supernova remnant from the \rosat\ HRI (top panel) and the
change in X-ray flux (middle panel) and expansion rate as a
function of radius (bottom panel) from a comparison of the two \rosat\
HRI observations. The error bars show the statistical uncertainty, 
while the boxes that surround each data point give an estimate of the 
systematic uncertainty.  The dashed line in the bottom panel is the 
global mean X-ray expansion rate.
}
\end{minipage}
\end{center}

\par

The published radio expansion parameter of Tycho's SNR,
$m=0.471\pm0.028$ (R97), corresponds to the expansion of the rim.
This value disagrees, by more than $3\, \sigma$, with the expansion
rate of the X-ray rim just derived.  Tycho's SNR thus joins the two
other youngest remnants of Galactic SN (Cas A and Kepler's SNR) in
showing considerably higher expansion rates in the X-ray compared to
the radio (see discussion and references in H99).  For Tycho's SNR it
is clear that the main difference between the X-ray and radio results
occurs at the remnant's outermost edge where the forward shock is
plowing into the ambient interstellar medium (ISM). The slower motion
of interior features (i.e., the reverse-shocked ejecta) is consistent
across the radio ($m\approx 0.44$; R97) and X-ray ($m\approx 0.45$)
bands.

\section{Discussion and Results}

Numerous authors (Chevalier 1982; Dwarkadas \& Chevalier 1998;
Truelove \& McKee 1999) have modeled the expansion rates of young
supernova remnants.  Dwarkadas \& Chevalier (1998) in particular
examined how different assumed density profiles for the SN ejecta
affect the evolution of the resulting remnant, assumed to be
interacting with a uniform density ISM.  They considered three
principal cases: a power law profile ($\rho \propto r^{-7}$), an
exponential profile, and a constant density profile.  The expansion
parameter uniquely defines the age and radius of the remnant,
conventionally expressed in scale-free variables.  The constants of
proportionality between the scaled values and the true physical radius
and age depend on the three independent dimensional parameters:
explosion energy $E$, ejecta mass $M_{\rm ej}$, and ambient density
$\rho_0$.  Thus, given the known size $R=2.8 (D/2.3\,\rm kpc)\,\rm pc$
and age $t=425\,\rm yr$ of Tycho's SNR it is possible to determine two
of the three dimensional parameters.  Here I assume that the ejecta
mass is $1.4\,M_\odot$ and solve for the other two quantities.

The power-law profile predicts a maximum expansion parameter of
$m=0.57$ which is too low to be consistent with the X-ray expansion
rate.  The other two model ejecta profiles can accommodate the high
rate observed for the forward shock; however, the inferred values of
$E$ and $\rho_0$ are quite different in the two scenarios. Compared to
the uniform density case, the expansion parameter for the exponential
profile model falls more rapidly with time, so that for a given value
of the expansion parameter the scaled age and radius are smaller for
the exponential profile.  The inferred dimensional parameters for the
exponential profile are $E_{51} = E/10^{51}\, {\rm erg} \approx (0.1 -
0.2)\, (M_{\rm ej}/1.4\, M_\odot)\,(D/2.3\,{\rm kpc})^2$ and $n_0 =
\rho_0/\mu_{\rm H} \approx (0.004 - 0.08)\, (M_{\rm ej}/1.4\,
M_\odot)\,(D/2.3\,{\rm kpc})^{-3}\,\rm cm^{-3}$ ($\mu_{\rm H}$ is the
mean mass per hydrogen atom), rather low values for Tycho's SNR.  On
the other hand, more appropriate values are obtained using the uniform
density ejecta model: $E_{51} \approx (0.4 - 0.5)\, (M_{\rm ej}/1.4\,
M_\odot)\,(D/2.3\,{\rm kpc})^2$ and $n_0 \approx (0.3 - 0.6)\, (M_{\rm
ej}/1.4\, M_\odot)\,(D/2.3\,{\rm kpc})^{-3}\,\rm cm^{-3}$.  These are
fully consistent with the values that Hamilton, Sarazin, \& Szymkowiak
(1986) found in their study of Tycho's X-ray spectrum in which they
also assumed a uniform density ejecta profile. Thus it appears that
both the expansion of the forward shock and the X-ray emission
properties of Tycho's SNR can be well explained with this simple
model. What about the slower motion of the reverse-shocked ejecta?

Truelove \& McKee (1999) have parameterized the evolution of both the
forward and reverse shocks in young SNRs for a number of cases,
including a uniform density ejecta model. For the range of scaled ages
that describe the expansion of the forward shock, the reverse shock
expansion parameter is $m_{\rm RS} = 0.51-0.63$.  This rate is in
reasonable agreement with the measured expansion parameter of the
interior portions of Tycho's SNR ($m = 0.45\pm0.04$). (Although
equating the motion of the ejecta to the reverse shock is not strictly
correct, it serves as a reasonable first approximation.)  An
acceptable joint fit ($\chi^2 = 2.7$ for 1 d.o.f.) to the measured
X-ray expansion parameters is obtained for best-fit values of
$m\approx0.64$ (forward) and $m\approx0.49$ (reverse).  The inferred
values of $E$ and $n_0$ in this case are similar to those quoted
above.

The forward shock expansion rate implies a shock velocity of
$4600\pm400\, (D/2.3\,{\rm kpc})\,\rm km\,s^{-1}$, which in turn
implies a mean post-shock temperature of $kT_S = {3\over 16}\, \mu m_p
v_S^2 = 25\pm4 (D/2.3\,{\rm kpc})^2\,\rm keV$ for a mean mass per
particle of $\mu =0.61$, which assumes a fully-ionized plasma with
10\% helium. This temperature is quite a bit higher than the estimate,
by Hwang, Hughes, \& Petre (1998), of the post-shock electron
temperature of the blast wave in Tycho's SNR, $kT_e \approx 4\,\rm
keV$.  This difference most likely arises from either the
non-equilibration of electron and ion temperatures at the shock front
or the partition of a significant fraction of the shock energy in
Tycho's SNR into relativistic particles as was recently found to be
the case in SNR E0102.2$-$7219 (Hughes, Rakowski, \& Decourchelle
2000).  Discriminating between these possibilities will be the focus
of a future article.

\acknowledgments

This research has made use of data obtained through the High Energy
Astrophysics Science Archive Research Center Online Service, provided
by the NASA/Goddard Space Flight Center.  Partial support was provided
by NASA grant NAG5-6420.

\end{document}